\documentclass[prd,aps,superscriptaddress,
amsfonts,amssymb,amsmath,showpacs,12pt]{revtex4-2}
\usepackage{bm}
\usepackage{amsfonts}
\usepackage{latexsym}
\usepackage[utf8]{inputenc}
\usepackage{graphicx}
\usepackage{amsmath}
\usepackage{palatino}
\usepackage{mathpazo}
\usepackage{textcomp}
\usepackage{booktabs}
\usepackage{dcolumn}
\usepackage{booktabs}
\usepackage{hyperref}
\hypersetup{colorlinks,citecolor=blue}
\usepackage{amsmath}
\usepackage{xcolor}
\usepackage{orcidlink}
\usepackage{caption}
\usepackage{commath}
\usepackage[justification=raggedright,singlelinecheck=false]{caption}

\allowdisplaybreaks

\newcommand{\oo}{\"{o}}

\newcommand{\tphi}{\tilde{\phi}}
\newcommand{\tet}{\tilde{e}{^a{}_{\mu}}}
\newcommand{\teti}{\tilde{e}{_{a}{}^{\mu}}}
\newcommand{\bw}{\bar{\omega}}

\begin{document}
	
	\title{Spontaneous scalarization of neutron stars in teleparallel gravity with derivative torsional coupling}
	
	\author{Youcef Kehal
	\orcidlink{0009-0004-9312-720X}}
	\email{youcef.kehal@univ-jijel.dz}
	
	\author{Khireddine Nouicer \orcidlink{0000-0001-6110-9833}}
	\email{khnouicer@univ-jijel.dz}
	
	\affiliation{Laboratory of Theoretical Physics and Department of Physics, Faculty of Exact and Computer Sciences, University of Jijel \\ B.P. 98, Ouled Aissa, Jijel 18000, Algeria}

\date{\today}

\begin{abstract}
We study neutron-star configurations in a teleparallel gravity model featuring a scalar field coupled to both matter and torsion. In the Einstein frame, the theory includes a derivative coupling between the scalar field and the torsion vector, together with a conformal matter coupling \(A(\phi)=\exp(\beta\phi^{2}/2)\). Static and slowly rotating neutron-star solutions are constructed for realistic equations of state, focusing on the APR and MS1 equations of state.
Scalarized solutions appear only within a finite range of central densities and correspond to localized deviations from the general-relativistic mass--radius and mass--central-density relations. The onset and extent of scalarization depend on the equation of state and on the strength of the derivative torsional interaction, which can either enhance or suppress scalarization relative to the general-relativistic scalarized branch. At high central densities, scalarization is quenched and the solutions approach the general-relativistic limit, remaining bounded even for strong torsional couplings. No scalarized solutions are found in the absence of matter coupling (\(\beta=0\)).
The normalized scalar charge follows trends consistent with the global mass relations, indicating an intermediate scalarized regime suppressed at high compactness. For slowly rotating stars, the moment of inertia depends systematically on the torsional coupling and the equation of state, with stiffer equations yielding larger values. These results highlight the potential of neutron-star radius and rotational measurements to test teleparallel scalarization scenarios.
\end{abstract}

\maketitle

\section{Introduction}\label{sec:intro}

The exploration of the strong-field regime has advanced significantly due to numerous gravitational wave (GW) detections from compact binary coalescences \cite{abbott2019gwtc, abbott2021gwtc, abbott2023gwtc} and observations of binary pulsars \cite{anderson2019binary,yordanov2024power}. In particular, neutron stars (NSs) act as natural laboratories for testing gravitational theories, allowing us to probe the accuracy of general relativity (GR) and delve deeper into potential extensions beyond it \cite{olmo2020stellar}.

In this light, scalar-tensor theories offer a promising way to extend GR, incorporating scalar fields as additional dynamical fields alongside the metric tensor \cite{quiros2019selected}. These theories can affect the properties of compact objects, with the process of spontaneous scalarization stands as a key example. This mechanism, initially proposed by the Damour and Esposito-Farèse (DEF) model \cite{Damour:1993,damour1996tensor},  consists of coupling the scalar field to the Ricci scalar or, equivalently to the trace of the energy-momentum tensor. Such coupling may develop a non-trivial structure only in the strong field regime of NSs, while remaining trivial and undetected in the weak field regime. Spontaneous scalarization can be understood as a phase transition, in which a linear tachyonic instability arises around a NS when its compactness exceeds a certain critical threshold \cite{doneva2024spontaneous}. This instability triggers the growth of the scalar field, as the effective mass becomes negative. Eventually, nonlinear effects quench the growth, causing the NS to acquire a scalar hair, which deviates from GR solutions inside the star, while still matching them at larger distances \cite{ventagli2022new}.

The original formulation of the DEF model focused solely on massless scalar fields but it is possible to incorporate massive or self-interacting scalar fields into the action \cite{chen2015spontaneous,ramazanouglu2016spontaneous}. The presence of potential leads to significant changes in the observational properties of NSs. On another front, it was shown that scalar Gauss-Bonnet (sGB) theories can undergo spontaneous scalarization in the context of both black holes (BHs) and NS \cite{silva2018spontaneous,doneva2018new,doneva2018neutron}, induced by the curvature of spacetime instead of matter \cite{blazquez2018radial}. Other studies determined the minimal
action that contains all the terms that can potentially trigger spontaneous
scalarization \cite{andreou2019spontaneous,ventagli2020onset} and thoroughly examined the impact of Ricci and Gauss-Bonnet couplings on scalarized NSs \cite{ventagli2021neutron}. 

Although, these developments have largely been framed within the realm of of Riemannian geometry, where gravity is described by the curvature of spacetime with a torsion-free and metric-compatible Levi-Civita connection.
Alternative formulations of gravity exist beyond this conventional approach. Notably, metric teleparallel theories of gravity provide a distinct geometric perspective in which the gravitational interaction is encoded in torsion instead of curvature \cite{bahamonde2021teleparallel}. Teleparallel gravity can be interpreted as a gauge theory of the translation group, formulated in terms tetrads and a flat spin connection \cite{aldrovandi2012teleparallel,krvsvsak2019teleparallel}. A dynamically equivalent formulation of GR is achieved when the gravitational action is constructed from quadratic contractions of the torsion tensor, with appropriately chosen coefficients, this theory is known as the Teleparallel Equivalent of GR (TEGR). Its Lagrangian differs from that of GR only by a total divergence term $ B $, ensuring that both theories yield the same field equations and physical predictions at astrophysical and cosmological scales. However, the equivalence no longer holds when the TEGR Lagrangian is promoted to an arbitrary function of the torsion scalar $ f(T) $ \cite{cai2016f} or by introducing scalar field into the action alongside its kinetic term $ X=-\frac{1}{2}g^{\mu \nu }\nabla_{\mu }\phi \nabla_{\nu }\phi $ and a derivative coupling with the torsion scalar $ Y=g^{\mu \nu }T^{\rho }{}_{\rho \mu }\nabla_{\nu }\phi$, giving rise to a broader class of models known as scalar-torsion theories $ \mathcal{L}(T,\phi,X,Y) $ \cite{hohmann2018scalar1,hohmann2018scalar2,hohmann2018scalar3}. These theories serve as a torsion-based counterpart to curvature-based scalar-tensor theories and significantly extend the landscape of several important subclasses, such as teleparallel dark energy models \cite{geng2011teleparallel}. On the other hand, the existence of hairy BH solutions in scalar-torsion theories was demonstrated in \cite{bahamonde2022scalarized}, and recent studies have begun exploring spontaneous scalarization of BHs in teleparallel sGB gravity \cite{bahamonde2023spontaneous,bahamonde2023distinctive}. However, the spontaneous scalarization of NS within the teleparallel framework remains largely unexplored and warrants further investigation.

To fill this gap, the main objective of this paper is to obtain the first teleparallel neutron star (NS) solutions exhibiting spontaneous scalarization. For this purpose, we allow the scalar field to be nonminimally coupled to the torsion scalar in the Jordan frame. After performing a conformal transformation, and with a specific choice of the parameter functions, the action can be rescaled to the Einstein frame, where a coupling function of the form $ \mathcal{C}(\phi)=\xi \phi^{2} $ multiplies the derivative coupling term. As a result, the scalar field is minimally coupled to gravity, while the matter fields are nonminimally coupled to the tetrad through the conformal function $ A(\phi)=\exp(\beta \phi^{2}/2) $. We investigate the properties of NSs in both static and rotating configurations by deriving the modified equilibrium equations under the assumption of a stationary, axisymmetric spacetime \cite{hartle1967slowly}. Despite persisting uncertainties in determining the internal EOS of NSs, we adopt two realistic candidates, APR and MS1, and model the stellar matter as a perfect fluid. Previous studies \cite{kehal2024neutron} have shown that nonminimal couplings with the torsion scalar, in combination with a Higgs-like potential, can lead to significant deviations from GR predictions. Extending this work, our results reveal the existence of two distinct branches of solutions: the standard GR branch and a nontrivial scalarized branch characterized by the emergence of scalar hair. Scalarized solutions are found for both positive and negative values of $ \xi $, but only within critical lower and upper thresholds beyond which scalarization ceases to exist. Within this bounded range, the star acquires a nonzero scalar charge, which acts as a source of scalar dipole radiation and could be probed by future gravitational-wave observations. We also explore how spontaneous scalarization modifies the moment of inertia in comparison with its behavior in GR.

This paper is organized as follows. In Sec. (\ref{sec:scator}), we introduce the general formalism of teleparallel gravity and present the field equations of the scalar torsion theory. In Sec. (\ref{sec:structure}), we derive the modified TOV equations governing the internal structure of neutron stars, following the Hartle–Thorne formalism, and specify the boundary conditions both at the center and at spatial infinity. Section (\ref{sec:results}) is devoted to the analysis of stellar properties, including the mass–radius relation, the evolution of the moment of inertia with mass, and other relevant features. Finally, in Sec. (\ref{sec:conclusion}), we summarize and discuss our main results.

\section{Scalar torsion theories}\label{sec:scator}
\subsection{Field equations}\label{subsec:action&feq}
The formulation of teleparallel geometry is built on the tetrad-spin connection pair. The tetrads $ e^{a}{}_{\mu } $ act as a soldering agents between the tangent space and general manifold, and are related to the metric tensor by
\begin{equation}\label{eq:metric}
	g_{\mu \nu }=\eta _{ab} e^a{}_{\mu}e^b{}_{\nu} \,,
\end{equation}
where $\eta _{ab}=\mathrm{diag}(-1,1,1,1)$ is the Minkowski metric. The spin connection 
$ \omega ^{a}{}_{b\mu } $ is assumed to be flat and metric compatible, and generated by local Lorentz transformation 
\begin{equation}\label{eq:spin-connection}
	\omega ^{a}{}_{b\mu }=\Lambda ^{a}{}_{c} \, \partial _{\mu }\Lambda _{b}{}^{c} \, .
\end{equation}
Thus, it represents a pure inertial gauge degree of freedom and one can always choose to work within a gauge under which it vanishes, commonly referred to as the Weitzenb\oo ck gauge \cite{aldrovandi2012teleparallel,bahamonde2021teleparallel,krvsvsak2019teleparallel}. According to teleparallel gravity, the gravitational strength is inherent in torsion tensor which is defined as the antisymmetric part of the Weitzenb\oo ck connection $\Gamma ^{\rho }{}_{\mu \nu }$:
\begin{equation}\label{eq:torsion_tensor}
	T^{\rho }{}_{\mu \nu }=-2\Gamma ^{\rho }{}_{[\mu \nu ]}=e_{a}{}^{\rho
	}\left( \partial _{\mu }e^{a}{}_{\nu }-\partial _{\nu }e^{a}{}_{\mu }+\omega
	^{a}{}_{b\mu }e^{b}{}_{\nu }-\omega ^{a}{}_{b\nu }e^{b}{}_{\mu }\right).
\end{equation}
We also define the superpotential tensor as: 
\begin{equation}\label{eq:super_potential}
	S_{\rho }{}^{\mu \nu }=K^{\mu\nu}{}_{\rho}-\delta _{\rho }^{\mu
	}T_{\sigma }{}^{\sigma \nu }+\delta _{\rho }^{\nu }T_{\sigma }{}^{\sigma \mu
	} \, ,
\end{equation}
where 
\begin{equation}\label{eq:contorsion}
 K^{\mu\nu}{}_{\rho}=	\frac{1}{2}\left(T^{\nu\mu}{}_{\rho}+T_{\rho}{}^{\mu\nu}-T^{\mu\nu}{}_{\rho}\right) 
\end{equation}
 is the contortion tensor. Finally, we are able to define the torsion scalar as the contraction  of the torsion tensor and its superpotential as:
\begin{equation}\label{eq:torsion_scalar}
	T=\frac{1}{2}S_{\rho }{}^{\mu \nu }T^{\rho }{}_{\mu \nu }.
\end{equation}

The TEGR action is constructed through a linear combination of the torsion scalar, which is related to the Ricci scalar $ R $ of GR (calculated using the Levi-Civita connection) by a total divergence term $ B $ as
\begin{equation}\label{eq:equivalence}
	R=-T+B, \quad  B=\frac{2}{e}\partial_{\mu }\left(eT_{\sigma}{}^{\sigma\mu}\right)=2\nabla_{\mu }T_{\nu}{}^{\nu\mu},
\end{equation}
where $e$ is the determinant of the tetrad $e^{a}{}_{\mu}$. The relation (\ref{eq:equivalence}) guarantees that the TEGR action produces the same field equations of GR, which is described by the following identity
\begin{equation}\label{eq:equivalence_identity}
	G_{\alpha \beta }=\overset{\circ }{\nabla }\vphantom{\nabla}_{\mu }S_{\beta
		\alpha }{}^{\mu }-\left(K^\mu{}_{\beta \nu}+T^\mu{}_{\beta \nu}\right)S_{\mu \alpha}{}^{\nu}+\frac{1}{2}g_{\alpha \beta}T \, ,
\end{equation}
where $ G_{\alpha \beta} $ is the Einstein tensor, and the right side represents the TEGR field equations. In a similar manner as the scalar tensor theories based on curvature \cite{quiros2019selected}, we upgrade the gravitational TEGR action to include scalar fields $ \phi $ as additional degree of freedom as follows
\begin{equation}\label{eq:action_general}
	\mathcal{S}_{g}=\int \mathcal{L}(T,\phi,X,Y) \, ,
\end{equation}
where we have define the kinetic term $ X $ and the derivative coupling $ Y $ of the scalar field as 
\begin{equation}\label{eq:kinetic&coupling}
	X=-\frac{1}{2}g^{\mu \nu }\nabla_{\mu }\phi \nabla_{\nu }\phi \quad , \quad Y=g^{\mu \nu }T^{\rho }{}_{\rho \mu }\nabla_{\nu }\phi \, .
\end{equation}
We derive the field equations of this theory in vacuum by varying the action (\ref{eq:action_general}) with respect to the tetrad field 
\begin{eqnarray}\label{eq:tetrad_feq_general}
	&&\mathcal{L}_{T} G_{\alpha\beta}+\frac{1}{2}g_{\alpha\beta}(\mathcal{L}-\mathcal{L}_{T}T)+S_{\beta\alpha}{}^{\nu}\nabla_{\nu}\mathcal{L}_{T}+\frac{1}{2}\mathcal{L}_{X} \phi_{\mu } \phi_{\nu} \\ \nonumber
	&&+\frac{1}{2}\nabla_{\sigma}(\mathcal{L}_{Y}\phi_{\rho})g^{\rho\sigma}g_{\alpha\beta}-\frac{1}{2}\nabla_{\beta}(\mathcal{L}_{Y}\phi_{\alpha})-\frac{1}{2}\mathcal{L}_{Y}\left(T_{(\alpha\beta)}{}^{\rho}\phi_{\rho}+\frac{1}{2}T^{\rho}{}_{\alpha\beta}\phi_{\rho}+T^{\sigma}{}_{\sigma\alpha}\phi_{\beta}\right)=0 \, ,
\end{eqnarray}
where we have used the notations $\mathcal{L}_{X}=\partial_{X} \mathcal{L}$ and
$\phi_{\mu}=\nabla_{\mu }\phi$.

The antisymmetric part of the above equation is the same as that obtained by varying the action (\ref{eq:action_general}) with respect to the spin connection
\begin{equation}\label{eq:spin_feq_general}
	2S_{[\beta\alpha]}{}^{\nu}\nabla_{\nu}\mathcal{L}_{T}+\nabla_{[\mu}\mathcal{L}_{Y}\phi_{\nu]}-\frac{3}{2}\mathcal{L}_{Y}T^{\rho}{}_{[\mu\nu}\phi_{\rho]}=0 \, .
\end{equation}
Throughout the paper, we will work in Weitzenb\oo ck gauge i.e. with symmetric tetrads compatible with vanishing spin connection \cite{hohmann2019modified} and satisfies the antisymmetric part of the field equation (\ref{eq:spin_feq_general}). Finally, the modified Klein Gordon equation for the scalar field reads as:
\begin{equation}\label{eq:scalar_feq_general}
	g^{\mu\nu}\nabla_{\mu}\left(\mathcal{L}_{Y}T^{\rho}{}_{\rho\nu}-\mathcal{L}_{X}\phi_{\nu}\right)=\mathcal{L}_{\phi} \, .
\end{equation}
\subsection{Conformal transformations}\label{subsec:conofrmal}

We consider a general class of scalar torsion theories in Jordan frame with $ \mathcal{A},\mathcal{B},\mathcal{C} $  free dimensionless functions of the scalar field $ \tphi $ \cite{hohmann2018scalar3}
\begin{equation}\label{eq:action_jordan}
	\tilde{\mathcal{S}}=\int d^{4}x \, \tilde{e} \left(-\tilde{\mathcal{A}}(\tphi)\tilde{T}-\tilde{\mathcal{B}}(\tphi) \tilde{g}^{\mu\nu} \nabla_{\mu }\tphi \, \nabla_{\nu }\tphi+\tilde{\mathcal{C}}(\tphi) \tilde{g}^{\mu\nu}\tilde{T}^{\rho}{}_{\rho\mu}\nabla_{\nu }\tphi \right)+\mathcal{S}_{m}\left(\Psi_{m},\tet \right) \, ,
\end{equation}
and where $ S_{m} $ stands for the action functional for the matter fields denoted by $ \Psi_{m} $ which are minimally coupled to the tetrad $ \tet $ (hereafter all quantities with a tilde are related to Jordan frame). We note that the presence of the derivative coupling term $ Y $ in the action (\ref{eq:action_jordan}) would lead to a nonminimal coupling to the boundary term $ \mathcal{D}(\tphi) B $ via an integration by parts, where $ \tilde{\mathcal{D}}(\tphi) \sim \int\tilde{\mathcal{C}}(\tphi) d\tphi$. 

We rescale the tetrad field by performing a conformal transformation
\begin{eqnarray}\label{eq:conformal}
	\tet=F^{-1}(\tphi) e^{a}{}_{\mu} \, \quad , \quad \teti =F(\tphi) e_{a}{}^{\mu} \, \quad , \quad \tilde{e}=F^{-4}(\tphi) e \, ,
\end{eqnarray}
where $ F(\tphi) $ is a conformal factor function that rescales the scalars (\ref{eq:torsion_scalar}) and (\ref{eq:kinetic&coupling}) as:
\begin{eqnarray}
	\tilde{T}=F^{2}\left(T-\frac{4F'}{F} Y+ \frac{12F'^{2}}{F^{2}}X\right) \, , \quad
	\tilde{X}=F^{2}\psi'^{2} X \, , \quad
	\tilde{Y}=F^2 \psi' \left(Y-\frac{6 F'}{F }X\right) \, ,
\end{eqnarray}
such that $ \psi(\tphi) $ is an arbitrary function of the scalar field. Taking the following parametrization 
\begin{eqnarray}
	\tilde{\mathcal{A}}(\tphi)&=&-\frac{\kappa}{2}F(\tphi)^{2} \, , \nonumber \\
	\tilde{\mathcal{B}}(\tphi)&=& \frac{F(\tphi)^{2}}{\psi'(\tphi)^{2}} \left[\frac{1}{2} +3 \frac{F'(\tphi)}{F(\tphi)}\left(\kappa \frac{F'(\tphi)}{F(\tphi)}+\mathcal{C}(\psi(\tphi))\right) \right] \, , \nonumber \\
	\tilde{\mathcal{C}}(\tphi)&=& \frac{F(\tphi)^{2}}{\psi'(\tphi)} \left[\mathcal{C}(\psi(\tphi))+2\kappa \frac{F'(\tphi)}{F(\tphi)}\right] \, , \nonumber
\end{eqnarray}
we reduce the action (\ref{eq:action_jordan}) in Einstein frame to the following suitable form
\begin{equation}\label{eq:action_einstein}
	S=\int d^{4}x \,e \left(\frac{\kappa}{2} T-\frac{1}{2} g^{\mu \nu}\nabla_{\mu }\phi \nabla_{\nu }\phi+\mathcal{C}(\phi)g^{\mu \nu}T^{\rho}{}_{\rho\mu}\nabla_{\nu }\phi\right)+S_{m}\left(\Psi_{m},	A(\phi)e^{a}{}_{\mu}\right) \, ,
\end{equation}
where $ \kappa=(8\pi G)^{-1} $ and we have redefined the scalar functions as
\begin{equation}
	\phi=\psi(\tphi) \qquad , \qquad A(\phi)=F^{-1}(\tphi) \, .
\end{equation}
In the absence of the coupling functions $ \mathcal{C}(\phi) $, the action (\ref{eq:action_einstein}) reduces to the standard action of a massless scalar field minimally coupled to gravity.

In this study, we investigate the implications of a quadratic form of the coupling function $ \mathcal{C}(\phi) $ 
\begin{equation}
	\mathcal{C}(\phi)=\xi \phi^{2} \, ,
\end{equation}
with $ \xi $ is an arbitrary constant. For this choice, the field equations (\ref{eq:tetrad_feq_general}) and (\ref{eq:scalar_feq_general}) simply take the form
\begin{eqnarray}\label{eq:tetrad_feq}
	G_{\alpha\beta}&=&\frac{1}{\kappa}\biggl[\xi \phi^{2}\, S_{\alpha\beta}{}^{\rho}\nabla_{\rho}\phi+\frac{1}{2}\left(1-4\xi \phi\right)g_{\alpha\beta} g^{\mu\nu}\nabla_{\mu }\phi \nabla_{\nu }\phi+\left(2\xi \phi-1\right)\nabla_{\alpha}\phi \nabla_{\beta }\phi \nonumber \\ && +\xi \phi^{2} \left(\nabla_{\beta }\nabla_{\alpha}\phi-g_{\alpha\beta}\Box \phi\right)-\Theta_{\alpha \beta }\biggl] \, , \\
	\Box \phi&=&2\xi \phi^{2} \, \nabla_{\mu }T_{\nu}{}^{\nu\mu} -\alpha(\phi) \, \Theta \, , \label{eq:phi_feq}
\end{eqnarray}
where $ \alpha(\phi)=d \ln A(\phi)/d\phi $ controls the coupling strength between the scalar field and matter \cite{harada1997stability} and $ \Theta_{\alpha \beta } $ is the energy-momentum tensor in Einstein frame with its trace $ \Theta=g^{\mu\nu}\Theta_{\mu \nu } $ related to the Jordan stress-energy tensor $ \tilde{\Theta}_{\mu\nu} $ and its trace  $ \tilde{\Theta}$ by 
\begin{eqnarray}\label{eq:conformal_fluid}
	\Theta_{\mu \nu }=A^{2}(\phi)\tilde{\Theta}_{\mu\nu} \,  , \,  \Theta^{\mu \nu}=A^{6}(\phi) \tilde{\Theta}^{\mu\nu} \, , \, \Theta=A^{4}(\phi)\tilde{\Theta}.
\end{eqnarray}
The contracted Bianchi identities for the energy-momentum tensor in the two frames satisfy
\begin{eqnarray}
	\nabla^{\beta}\Theta_{\alpha\beta}&=&\xi \phi^{2}\, \nabla_{\mu}S_{\alpha}{}^{\mu\nu}\phi_{\nu} -\phi_{\alpha}\Box \phi \, ,\\
	\nabla^{\beta}\tilde{\Theta}_{\alpha\beta}&=&0 \, .
\end{eqnarray}
This results in particles moving on geodesics of the metric in Jordan frame while, they are following trajectories that are influenced by the scalar field in Einstein frame.

\section{Structure equations}\label{sec:structure}
To derive the structural equations for slowly rotating isotropic NS, we follow the Hartle-Thorne formalism  \cite{hartle1967slowly} adopting a stationary axisymmetric spacetime to the first order of the angular velocity $ \Omega $. This latter is assumed to be uniform and small enough not to perturb the physical properties of the star. The line element is given by
\begin{equation}\label{eq:metric_rotation}
	ds^{2}=-f(r)^{2}dt^{2}+h(r)^{2}dr^{2}+r^{2}(d\theta ^{2}+\sin^{2}\theta d\varphi
	^{2})-2\omega (r,\theta )r^{2}\sin^{2}\theta dtd\varphi +\mathcal{O}(\Omega
	^{2}) \, ,
\end{equation}
where $\omega (r,\theta )=d\varphi/dt $ accounts for frame dragging effect  proportional to the angular velocity $ \Omega $. Through the conformal transformation $ d\tilde{s}^{2}=A(\phi)^{2}ds^{2} $, the distance and metric functions in the Jordan frame are related to those in the Einstein frame by
\begin{eqnarray}\label{eq:metric-conformal}
	\tilde{r}=A(\phi) r \, , \, \tilde{f}(\tilde{r})=A(\phi) f(r) \, , \,
	\tilde{h}(\tilde{r})= \left(1+\alpha(\phi)r\frac{d\phi}{dr}\, \right)^{-1}  h(r) \, , \, \tilde{\omega} (\tilde{r},\theta )=\omega (r,\theta ).
\end{eqnarray}
One can recover the metric (\ref{eq:metric_rotation}) by using the regular branch of the axially symmetric Weitzenb\oo ck tetrads \cite{bahamonde2021exploring}
\begin{equation}
	e^{a}{}_{\mu}=\left( 
	\begin{array}{cccc}
		f(r) & 0 & 0 & \frac{r^{2}\sin ^{2} \theta  \omega (r,\theta )  }{f(r)}\\ 
		0 & h(r)\sin \theta \cos \varphi  & r\cos \theta \cos \varphi & 
		-r\sin \theta \sin \varphi \\ 
		0 & h(r)\sin \theta \sin \varphi  & r\cos \theta \sin \varphi  & 
		r\sin \theta \cos \varphi \\ 
		0 & h(r)\cos \theta & -r\sin \theta  & 0
	\end{array}
	\right).
\end{equation}
This choice of the tetrad is compatible with  a zero spin connection, i.e. the Weitzenb\oo ck tetrad that satisfies the antisymmetric part of the field equations(\ref{eq:spin_feq_general}). Thus, the torsion scalar and the boundary term can be calculated through  (\ref{eq:torsion_scalar})-(\ref{eq:equivalence}) and are:
\begin{eqnarray}
	T&=&\frac{2 (h-1) \left(2 r f'-f h +f \right)}{r^2 f h^2} \, , \\
	B&=&\frac{2 r \left[f' \left(r h'+2h (h-2) \right)-h r f''\right]+4 f \left(r h'+h (h-1) \right)}{r^2 f h^3} \, .
\end{eqnarray}
By modeling the stellar matter inside the star as a perfect fluid we write the stress-energy tensor in Jordan frame as
\begin{equation}
	\tilde{\Theta}_{\mu \nu }=\left(\rho+P\right) \tilde{u}_{\mu}\tilde{u}_{\nu} + \tilde{g}_{\mu\nu} P \, ,
\end{equation}
where we omit the tilde notation for pressure and density, as we will focus solely on these quantities within the Jordan frame. Note that the relations between the stress energy tensor in Jordan and Einstein frames  are given by (\ref{eq:conformal_fluid}), and the fluid velocity in both frames are related as follows $ u^{\mu}=A(\phi)\tilde{u}^{\mu} $, we obtain that:
\begin{equation} 
	u^{\mu}=(u^{0},0,0,\Omega u^{0}) \quad \quad 
	\text{with}  \quad
	u^{0}=\left[-\left(g_{tt}+2\Omega g_{t\varphi}+\Omega^{2}g_{\varphi\varphi}\right)\right]^{-1/2} \, .
\end{equation}
Thus, the angular velocity $ \Omega=u^{\varphi}/u^{t} $ is the same for both frames and the Einstein frame stress-energy tensor for the perfect fluid can be written as \cite{pani2014slowly}
\begin{equation}
	\Theta_{\mu \nu }=A^{4}(\phi)g_{\mu \rho}g_{\nu \sigma}\left[(\rho+P)u^{\rho}u^{\sigma}+P g^{\rho \sigma}\right].
\end{equation}
It follows that:
\begin{eqnarray}
	\Theta^{\mu}_{\nu}=A^{4}(\phi) \text{diag}(-\rho,P,P,P), \quad \Theta=A^{4}(\phi)(3P-\rho) \, ,
\end{eqnarray}
and 
\begin{eqnarray}
	\Theta^{\varphi}_{t}=\frac{r^{2}\sin(\theta)^{2}}{f^2}A^{4}(\phi) (\rho+P)\Omega \bw \, ,
\end{eqnarray}
where we define the function $ \omega(r,\theta)=\Omega (1-\bw(r,\theta)) $. We can proceed now to solve the Einstein frame
field equations (\ref{eq:tetrad_feq}) and  (\ref{eq:phi_feq}) as well as the contracted Bianchi identity
\begin{eqnarray}
	\frac{f'}{f}&=&\frac{-1+h^2+4\pi r \left(r\phi'^2-4\xi \phi' \phi^2+2rA^{4} h^2 P\right)}{2r \left(1+4\pi \xi r \phi' \phi^2\right)} \, , \label{eq:tov1}\\
	\frac{h'}{h}&=&\frac{1+h^2\left(8\pi r^2 A^4  \rho-1\right)+4\pi r^2 \phi'^2+H_{1}\xi +H_{2}\xi^2+H_{3}\xi^3+H_{4}\xi^4}{2r\left(1+12\pi \xi^2 \phi^4\right)\left(1+4\pi  \xi r \phi^2 \phi'\right)} \, , \label{eq:tov2}\\
	\phi''&=&\frac{r h^2  \phi ' \left(4 \pi r^2 A^4 (\rho -P)-1\right)+r^2 \alpha A^4 h^2  (3 P-\rho )-r \phi '+\Phi_{1}\xi+\Phi_{2}\xi^2+\Phi_{3}\xi^3}{r^2(1+12\pi \xi^{2}\phi^{4})}  \label{eq:tov3} \, , \\
	P'&=&-(\rho+P)\left[\alpha \phi'+\frac{h^2 \left(8 \pi  r^2 A^4 P+1\right)+4 \pi  r \phi ' \left(r \phi '-4 \xi  \phi
		^2\right)-1}{2 r \left(4 \pi  \xi  r \phi^2 \phi '+1\right)}\right] \, , \label{eq:tov4}
\end{eqnarray}
where the coefficients $ H_{i} $ and $ \Phi_{i} $ are given in the Appendix. 

We define the Arnowitt-Deser-Misner (ADM) mass $ M $ in Einstein frame as 
\begin{eqnarray}
	M=\lim_{r \rightarrow \infty} \mathcal{M}(r)=\frac{r}{2}(1-h^{-2}) \, .
\end{eqnarray}

Clearly, in the absence of scalar fields (for  $ A(0)=1 $ and $ \xi=0 $ )   the equation (\ref{eq:tov4}) reduces to GR TOV equation
\begin{eqnarray}\label{eq:tov-gr}
	P'=-(\rho+P)\frac{4\pi r^3 P+M}{r(r-2M)}\, .
\end{eqnarray}
Taking into account the field equation at the linear order of $ \Omega $ and expanding the function $ \bw(r,\theta) $ on the basis of the Legendre polynomial $ P_{l} $ as follows
\begin{equation}
	\bw(r,\theta)=\sum_{l=1}^{\infty}\bw_{l}(r)\left(-\frac{1}{\sin(\theta)}\frac{dP_{l}}{d\theta}\right) \, ,
\end{equation}
we can write the radial differential equation for $ \bw $ as
\begin{eqnarray}
	\partial_{r}\left(\frac{r^{4}\bw'_{l}}{fh}\right)+\frac{r^2}{f}\left\{h\left[l\left(l+1\right)-2\right]-\frac{2\xi r}{\kappa}\phi^{2}\phi'\right\}\bw_{l}+\frac{2\xi r^{3}}{\kappa f}\phi^{2}\phi'=\frac{4 A^{4}r^{4}}{\kappa}\frac{h}{f}\left(\rho+P\right)\bw_{l}.\nonumber \\	
\end{eqnarray}
The requirement of asymptotic flatness implies that $ \bw_{l}=0 $ for $ l \ge 2 $, thus $ \bw $ is a function of $ r $ only and $ l=1 $. Using Eqs. (\ref{eq:tov1})-(\ref{eq:tov3}), we obtain
\begin{equation}\label{eq:tov5}
	\bw''=\frac{-16\pi \xi r h \phi' \phi^{2} (1+12\pi \xi^{2}\phi^{4})(1+4\pi\xi r \phi' \phi^{2})+B \bw + C \bw'}{r^{2}(1+12\pi \xi^{2}\phi^{4})(2+8\pi \xi r \phi' \phi^{2})} \, ,
\end{equation}
with $ B $ given by
\begin{eqnarray}
	B(r)=16\pi r h (1+12\pi \xi^{2}\phi^{4})(1+4\pi \xi r \phi' \phi^{2})\left[2r A^{4}h(\rho+P)+\xi \phi' \phi^{2}\right] \, ,
\end{eqnarray}
and the coefficient $ C $ is expressed as a series of power of $r$ as $ C=\sum_{i=1}^{4} \bar{C}_{i} r^{i}$ and the coefficients $ \bar{C}_{i} $ are listed in Appendix. The system of equations (\ref{eq:tov1})–(\ref{eq:tov4}) and (\ref{eq:tov5}) governs the structure of both the interior and exterior of a slowly rotating star. To render this system fully determinate, it must be closed by specifying an equation of state, a functional relationship between the pressure $ P $
and the energy density $ \rho $. In this context, we assume a barotropic equation of state $ P=P(\rho) $. Additionally, appropriate boundary conditions must be imposed at the center of the star and at spatial infinity to ensure physically meaningful solutions.

\subsection{Expansion at the center}
Near the center of the star, we can carry out an analytic expansion in the form of
\begin{equation}\label{eq:initial_center}
	\epsilon=\sum_{n=0}^{\infty}\epsilon_{n}r^{n} \, ,
\end{equation}
for the metric functions, the scalar field, and pressure. The regularity at the center lead to $\epsilon_{1}=0$ for al the quantities. At this stage, The equilibrium equations can be expressed in terms of two parameters, which can be freely determined: the central density $ \rho_{0} $ (the central pressure $ P_{0} $ is determined directly once $ \rho_{0} $ is fixed) and the value of the scalar field at the center $ \phi_{0} $. Thus, all higher order coefficients can be derived as functions of these parameters
\begin{eqnarray}\label{eq:initial}
	f_{2} & \equiv & \frac{2\pi A_{0}^{4}\left[ 3P_{0}+\rho_{0}+8\pi \xi^{2}\phi_{0}^{4}(3P_{0}+2 \rho_{0})+\xi \alpha_{0}\phi_{0}^{2}\left(3P_{0}-\rho_{0}\right)\right]}{3(1+12\pi \xi^{2}\phi_{0}^{4})} \, , \\
	h_{2}& \equiv & \frac{2\pi A_{0}^{4}\left[ 3P_{0}+\rho_{0}+8\pi \xi^{2}\phi_{0}^{4}(3P_{0}+2 \rho_{0})+\xi \alpha_{0}\phi_{0}^{2}\left(3P_{0}-\rho_{0}\right)\right]}{3(1+12\pi \xi^{2}\phi_{0}^{4})} \, , \\
	P_{2} & \equiv & -\frac{A_{0}^{4}(\rho_{0}+P_{0})\left[4\pi (3P_{0}+\rho_{0}+8\pi \xi^{2}\phi_{0}^{4}(3P_{0}+2\rho_{0}))-\alpha_{0}(3P_{0}-\rho_{0})(\alpha_{0}-8\pi\xi\phi_{0}^{2})\right]}{6(1+12\pi \xi^{2}\phi^{4}_{0})},\nonumber \\
	 \\
	\phi_{2}& \equiv & \frac{A_{0}^{4}(3P_{0}-\rho_{0})(4\pi \xi \phi_{0}^{2}-\alpha_{0})}{6(1+12\pi \xi^{2}\phi^{4}_{0})} \, ,
\end{eqnarray}
where $ A_{0}=A(\phi_{0})$ and $ \alpha_{0}=A'(\phi_{0})/A(\phi_{0}) $.  On the other hand, we set $ h_{0}=1 $ to avoid singularity and $ f_{0}=1 $ by time rescaling at the center. A similar analysis can be done for rotation case in terms of $ \bw_{0} $
\begin{align}
	\bw_{2} \equiv \frac{8\pi A_{0}^{4}\left[3\bw_{0}(\rho_{0}+P_{0})-\xi \alpha_{0}\left(\bw_{0}-1\right)\left(3P_{0}-\rho_{0}\right)\phi_{0}^{2}+4\pi\xi^{2} \left(3P_{0}\left(4\bw_{0}-1\right)+\rho_{0}\left(8\bw_{0}+1\right)\right)\phi_{0}^{4}\right]}{15(1+12\pi \xi^{2}\phi_{0}^{4})} \, ,
\end{align}
Here, we observe that all quantities are multiplied by the conformal factor, even in the presence of the coupling $\xi$. This implies that removing the coupling between matter and the tetrad would eliminate the scalarization phenomenon. In the expansions derived above, coefficients beyond order 4 are omitted due to their lengthy expressions. We note that the central scalar field $ \phi_{0} $ and $ \bw_{0} $ are to be determined using a shooting method to satisfy asymptotic flatness at infinity.

Let us now derive the conditions under which the spharically symmetric NS solutions  are stable demanding that $P''(r)<0$. We have tow cases dependong on teh sign of the first derivative of the scalar field. In fact we obtain for $ (\phi,\,\xi) \in \mathbb{R}$:
\begin{enumerate}
\item $P''(r)<0$ and $\phi'(r)<0$: 
\begin{eqnarray}
&\text{If}&\,\,0<\rho_0 <3 P_0,\qquad \alpha_0<-4\pi\xi \phi_0^2\\
&\text{If}&\,\,\rho_0>3P_0,\qquad 
-4\pi\xi \phi_0^2<\alpha_0<2\left(\frac{\pi(3P_0+\rho_0)
+8\pi^2\xi^2\phi_0^4(3P_0+2\rho_0)}{\rho_0-3P_0}\right)^{1/2}\nonumber\\
\end{eqnarray}
\item $P''(r)<0$ and $\phi'(r)>0$: 
\begin{eqnarray}
&\text{If}&\,\,0<\rho_0 <3 P_0,\qquad \alpha_0>-4\pi\xi \phi_0^2\\
&\text{If}&\,\,\rho_0>3P_0,\qquad 
-2\left(\frac{\pi(3P_0+\rho_0)
+8\pi^2\xi^2\phi_0^4(3P_0+2\rho_0)}{\rho_0-3P_0}\right)^{1/2}  <\alpha_0<-4\pi\xi \phi_0^2\nonumber\\
\end{eqnarray}
\end{enumerate}

\subsection{Asymptotic Expansion at infinity}
At large distances, both the energy density and pressure vanish. The asymptotic behavior of the metric functions and the scalar field can be expressed as a series in inverse powers of $r$:
\begin{eqnarray}
	f=\sum_{n=0}^{\infty}\frac{f^{\infty}_{n}}{r^{n}} \, , \qquad
	h=\sum_{n=0}^{\infty}\frac{h^{\infty}_{n}}{r^{n}} \, , \qquad 
	\phi=\sum_{n=0}^{\infty}\frac{\phi^{\infty}_{n}}{r^{n}} \, ,
\end{eqnarray}
where $ f^{\infty}_{n} $, $ h^{\infty}_{n} $, and $ \phi^{\infty}_{n} $ are constants. Asymptotic flatness requires that $ f^{\infty}_{0} = h^{\infty}_{0} = 1 $, and for simplicity, we impose that the asymptotic value of the scalar field vanishes, i.e., $ \phi^{\infty}_{0} = 0 $. By substituting these expansions into the field equations, we can solve them order by order to obtain the asymptotic expansion up to fourth order as follows:
\begin{eqnarray}\label{eq:infinity}
	f&=&1-\frac{M}{r}-\frac{M^{2}}{r^{2}}-\frac{\left(4M^{3}-2\pi M Q^{2}+4\pi \xi Q^{2}\right)}{3r^{3}}\nonumber\\
	&-&\frac{\left(6M^{4}+8\pi M^{2}Q^{2}+40\pi \xi  M Q^{3}\right)}{3r^{4}}+{\cal{O}}\left(1/r^5\right),\nonumber \\
	 \\
	h&=&1+\frac{M}{r}+\frac{3M^{2}-4\pi Q^{2}}{2r^{2}}+\frac{5M^{2}-16\pi M Q^{2}-8\pi \xi Q^{3}}{2r^{3}}\nonumber\\
	&+&\frac{35M^{4}-\frac{568\pi}{3}M^{2}Q^{2}+48\pi^2 Q^4-\frac{448}{3}\pi\xi M Q^{3}}{8r^{4}}+{\cal{O}}\left(1/r^5\right),  \\
	\phi&=&\frac{Q}{r}+\frac{M Q}{r^{2}}+\frac{4M^{2}Q-2\pi Q^3}{3r^{3}}+\frac{2M^{3}Q-\frac{8 \pi}{3}M Q^{2}-\frac{\xi}{4}\left(\frac{M^{2}Q^{3}}{3}+4\pi Q^{4}\right)}{r^{4}} +{\cal{O}}\left(1/r^5\right),\nonumber\\
\end{eqnarray}
where $ M $ and $ Q=\phi^{\infty}_{1} $ correspond to the star's mass and scalar field charge (monopole) in Einstein frame, respectively. These quantities are related to their counterparts in the Jordan frame through
\begin{eqnarray}
	\tilde{M}&=&A(\phi)\left[M-\alpha(\phi)r^{2}\frac{d\phi}{dr}\left(1-\frac{2M}{r}\right)\left(1+\frac{\alpha(\phi)r}{2} \frac{d\phi}{dr}\right)\right] \, , \\
	\tilde{Q}&=&A(\phi) Q (1-\alpha \frac{Q}{r})\frac{d\tilde{\phi}}{d\phi}.
\end{eqnarray}
Furthermore, The asymptotic expansion of $ \bw $ is written as follows
\begin{equation}
	\bw=\Omega-\frac{2J}{r^{3}}+\frac{12\pi J Q^{2}}{5 r^{5}} +{\cal{O}}\left(1/r^6\right)\, ,
\end{equation}
where $ J $ is the angular momentum of the star which is related to the moment of inertia as
\begin{equation}
	I=\frac{J}{\Omega}.
\end{equation}
Note that the higher-order terms in the expansion of $\bw$ involve the coupling parameter $\xi$, thereby differing from the corresponding asymptotic expansion in GR. As expected, all expressions of GR are recovered in the limit $\xi \to 0$ \cite{pani2014slowly}.

In the numerical implementation, we employ a double shooting method to determine the moment of inertia of the star. Or it can be calculated directly by the  formula
\begin{equation}
	I=\frac{16\pi}{3}\int_{0}^{r_{s}}A^{4}r^{4}\frac{h}{f}(\rho+P)\bw dr+\frac{8\pi\xi}{3} \int_{0}^{r_{s}}\frac{r^2}{f}(1-\bw)\phi^{2}\phi' dr .
\end{equation}

\section{Results}\label{sec:results}

The numerical integration of the system of equations is determined by the value of the energy density at the center $ \rho_{c} $ with the central pressure $ P_{c} $ being determined directly by the EOS. We also set the asymptotic value of the scalar field at infinity very small, $ \phi^{\infty}_{0}=10^{-3} $ \cite{pani2014slowly}. The TOV equations are then integrated numerically from the star’s center outward, using the initial conditions (\ref{eq:initial_center}) up to 6th order, until the pressure drops to zero, defining the star’s surface $ P(r_{s})=0 $. Additionally, we integrate from the surface to infinity and ensure that the metric functions and scalar field smoothly satisfy the asymptotic boundary conditions at large distances (\ref{eq:infinity}). Throughout this manuscript, we assume 2 EOS: APR and MS1 which are taken to be barotropic $ P(\rho) $, suitable for cold NSs whose temperatures lie well below the Fermi temperature, making their pressure effectively a function of density alone \cite{silva2024neutron}. 
\subsection{Damour-Esposito-Farese model}
In this study we adopt the Damour-Esposito-Farese model conformal factor between the matter and the scalar field \cite{Damour:1993}
\begin{eqnarray}\label{eq:conformal-factor}
	A(\phi)=e^{\frac{\beta}{2} \phi^{2}},
\end{eqnarray}
where $\phi=0$ is also a solution of our model. Let us now consider a small perturbation $ \delta \phi $ around a constant scalar field background $ \phi_{0} $. At linear level, Eq. (\ref{eq:phi_feq}) becomes
\begin{equation}
	\left (\Box  - m^{2}_{eff}\right )\delta \phi=0,
\end{equation}
where the the square of the scalar field
effective mass is defined as
\begin{eqnarray}
	m_{eff}^2=4\xi\phi\nabla_{\mu }T_{\nu}{}^{\nu\mu}-\beta\Theta.
\end{eqnarray}
A negative value of the square of the effective mass signals the onset of a tachyonic instability, whereby perturbations with sufficiently small wave numbers undergo exponential growth, leading to spontaneous scalarization \cite{doneva2024spontaneous}.
The term proportional to the matter coupling constant $\beta$ plays a role analogous to that encountered in scalar--tensor theories formulated in the Einstein frame. Inside neutron stars, the trace of the matter energy--momentum tensor $\Theta$ is typically negative, so that for $\beta<0$ the contribution $-\beta\Theta$ can render $m_{\rm eff}^2$ negative. This induces a tachyonic instability of the scalar perturbations, leading to the spontaneous growth of the scalar field and the appearance of scalarized solutions, in close analogy with the general-relativistic scalarization mechanism \cite{Harada:1997,Damour:1993}. 
By contrast, the first term proportional to the derivative torsional coupling $\xi$ has a distinct geometrical origin. It depends on derivatives of the torsion vector and is therefore sensitive to gradients of the gravitational field rather than directly to the local matter density. Its contribution to $m_{\rm eff}^2$ can be either positive or negative, depending on the sign of $\xi$, the background scalar-field configuration, and the internal structure of the neutron star.
As a result, the torsional term can either enhance or counteract the tachyonic instability driven by the matter coupling. When it contributes positively to $m_{\rm eff}^2$, it acts as a stabilizing effect, suppressing scalarization and eventually restoring the general-relativistic branch at sufficiently high central densities. Conversely, when its contribution is negative, it can strengthen the instability and enlarge the region of parameter space where scalarized solutions exist. The competition between these two terms naturally leads to a bounded scalarized regime, in which scalarization occurs only over a finite range of central densities and does not grow indefinitely as the torsional coupling is increased.
This interplay between the matter-induced and torsion-induced contributions to the effective mass provides a clear physical explanation for the emergence of an intermediate scalarized phase and its quenching at high compactness, highlighting a key qualitative difference between teleparallel scalarization and its scalar--tensor counterpart.

In the numerical integration, we adopt the values $ \beta = -4\pi \times 6 $ \cite{silva2024neutron} and $ \beta = -4\pi \times 4.8 $, both of which are sufficient to trigger spontaneous scalarization, regardless of the EoS. Although the work of \cite{Damour:1993} estimated that $ \beta \le -4\pi \times 4 $ is sufficient, this threshold has since been refined through multiple binary pulsar observations, suggesting that
$ \beta \le -4\pi\times 4.35 $ is more accurate \cite{harada1997stability}.

\begin{figure}[ht]
	\includegraphics[height=4cm,width=0.32\textwidth]{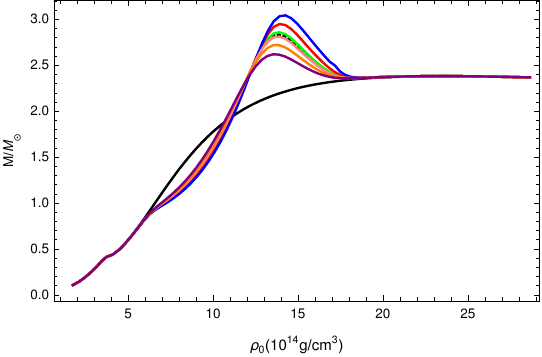}
	\includegraphics[height=4cm,width=0.32\textwidth]{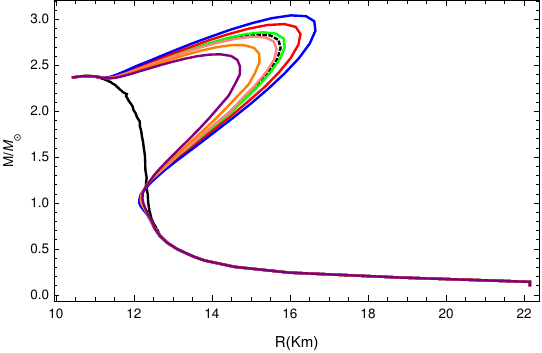}
	\includegraphics[height=4cm,width=0.32\textwidth]{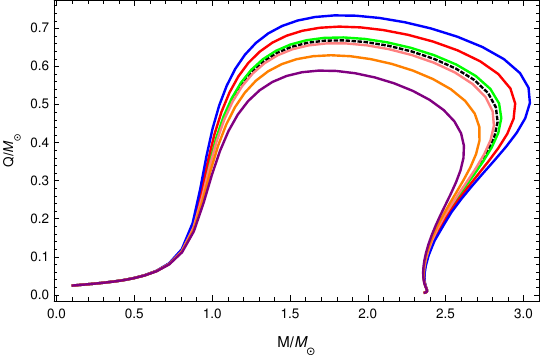}
	\caption{Properties of NS for $\beta =-4\pi\times 4.8$ using the APR EOS.	We show the mass $M$ vs the central energy density $\rho_{c}$ (left panel), the mass $M$ vs the star radius $R$ (middle panel), and the normalized scalar charge $Q$ vs the mass $M$ (right panel). The solid black curves correspond to the GR solutions while the black dashed curvess correspond to the uncoupled case $ \xi=0 $. The colored solid curves correspond to the values: $ \xi=0.5 $ (green), $ \xi=2.5 $ (red) and $ \xi=5 $ (blue) and $ \xi=-0.5 $  (pink), $ \xi=-2.5 $ (orange), $ \xi=-5 $ (purple).}
	\label{fig:plots_apr}
\end{figure}

\begin{figure}[ht]
	\includegraphics[height=4cm,width=0.32\textwidth]{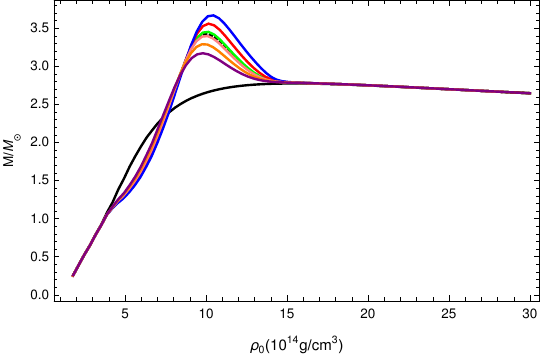}
	\includegraphics[height=4cm,width=0.32\textwidth]{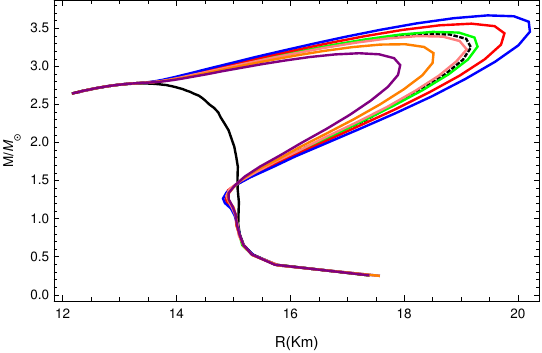}
	\includegraphics[height=4cm,width=0.32\textwidth]{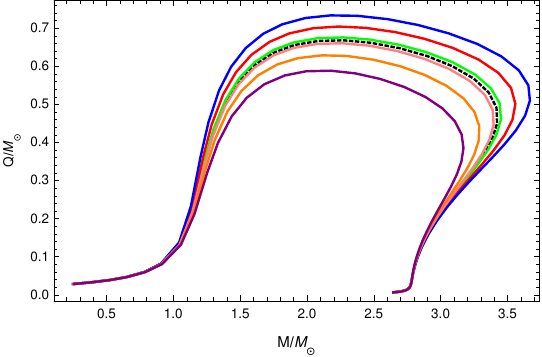}
	\caption{Properties of NS for the MSI EOS and  $\beta =-4\pi\times 6$ We use the same parameters and colors as in Fig. \ref{fig:plots_apr}.}
	\label{fig:plots_ms1}
\end{figure}

We have plotted several key properties of NS for two different EOS in figures \ref{fig:plots_apr} and \ref{fig:plots_ms1} : the mass-central density $ \rho_{c} $ and the mass-radius diagrams, and the normalized charge as a function of mass. The analysis reveals the emergence of multiple solution branches: the standard GR branch and a scalarized branch characterized by a nontrivial scalar field. 

\subsection{Teleparallel effects and bounded scalarization}

For both the APR ($\beta=-4\pi\times 4.8$) and MS1 ($\beta=-4\pi\times 6$) EOS, the $M$--$\rho_c$ and $M$--$R$ relations shown in the right and middle panels of Figs.~\ref{fig:plots_apr} and \ref{fig:plots_ms1} exhibit a scalarization-induced bump whose location and amplitude depend on the stiffness of the EOS and on the teleparallel coupling parameter $\xi$. In the limit $\xi=0$, the solutions reduce to the standard scalarized branch familiar from GR-like scalar--tensor theories. When the derivative torsion interaction is included, a clear ordering of the scalarized sequences emerges: for positive values of $\xi$ the scalarized branch lies above the GR scalarized sequence, whereas for negative $\xi$ it lies below it. This hierarchy indicates that the teleparallel coupling enhances or suppresses the effective stellar support for $\xi>0$ and $\xi<0$, respectively.

The bump corresponds to an intermediate scalarized regime along a single continuous equilibrium sequence. As the central density increases, the negative matter coupling $A(\phi)=e^{\beta\phi^2/2}$ activates the scalar field and drives a departure from the GR branch. In this regime, the derivative torsion interaction $\xi \phi^2 \nabla_\mu \phi\, T^\mu$ modifies the hydrostatic balance, allowing for enhanced gravitational masses when $\xi>0$ and reduced deviations from GR when $\xi<0$. At higher central densities, scalarization is progressively quenched and the mass sequence approaches the GR curve again. The descalarization slope depends strongly on $\xi$: the suppression of scalarization beyond the maximum mass is most pronounced for $\xi>0$, milder for $\xi=0$, and weakest for $\xi<0$, reflecting the increasing energetic cost of large scalar gradients induced by a positive teleparallel coupling.

Scalarization does not grow indefinitely with increasing or decreasing $\xi$. Instead, we identify two threshold values, $\xi_{\min}<0$ and $\xi_{\max}>0$, beyond which the scalarized solutions saturate. For $\xi>\xi_{\max}$, the scalarized branch becomes effectively indistinguishable from the branch obtained at $\xi=\xi_{\max}$, while for $\xi<\xi_{\min}$ the solutions remain stuck to the $\xi_{\min}$ branch. Although these threshold values cannot be determined analytically due to the nonlinear coupling between torsion, scalar gradients, and matter, they can be estimated numerically by identifying the values of $\xi$ beyond which further variation produces no significant change in the mass--radius relation or in scalar observables.

This interpretation is further supported by the behavior of the normalized scalar charge $Q/M$ as a function of the gravitational mass as shown in the right panels of figures \ref{fig:plots_apr} and \ref{fig:plots_ms1}. The scalar charge exhibits the same qualitative trends as the mass diagrams: it grows rapidly in the intermediate scalarized regime, reaches a maximum in correspondence with the mass enhancement region, and is systematically larger for $\xi>0$ and smaller for $\xi<0$ compared to the GR scalarized branch. At larger masses, $Q/M$ decreases and approaches zero, signaling the descalarization of highly compact configurations and the recovery of GR-like behavior. The decay of the scalar charge is steepest for $\xi>0$, milder for $\xi=0$, and weakest for $\xi<0$, and it likewise saturates for $|\xi|>\{\xi_{\min},\xi_{\max}\}$, confirming that the teleparallel derivative interaction regulates the strength of scalarization without inducing unbounded growth of scalar hair.

Finally, we note that for vanishing matter coupling ($\beta=0$), scalarized solutions disappear even when $\xi\neq 0$. This demonstrates that spontaneous scalarization remains fundamentally driven by the matter--scalar coupling, while the teleparallel derivative interaction controls the magnitude, saturation, and termination of the scalarized regime.
\subsection{Dependence on the matter coupling $\beta$.}
\begin{figure}[h]
	\centering
	\includegraphics[width=0.45\textwidth]{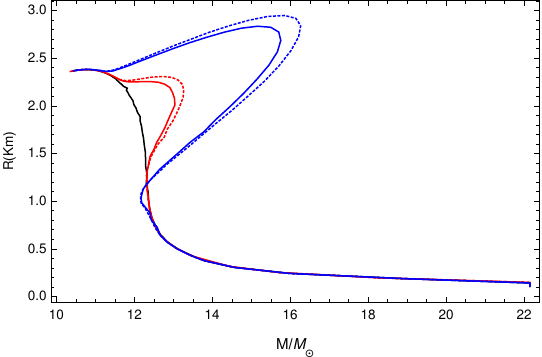}
	\includegraphics[width=0.46\textwidth]{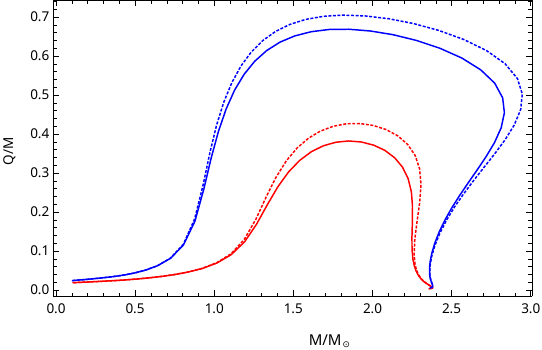}
	\caption{Effect of varying $ \beta $ on the scalarized branch solutions on mass-radius and the charge-mass relations for the APR EOS. The black solide curves is the standard GR solutions while the red curves are the scalarized solutions for $ \beta=-4\pi\times 4.8 $ (solid for $ \xi=0 $ and dashed for $ \xi=2.5 $) and blue curves for $ \beta=-4\pi\times 6 $ (solid for $ \xi=0 $ and dashed for $ \xi=2.5 $). }
	\label{fig:MRbeta}
\end{figure}

We have also investigated the impact of the matter--scalar coupling parameter $\beta$ and the derivative torsional coupling $\xi$ on the strength of scalarization by considering the APR equation of state for two representative values, $\beta = -4\pi \times 4.8$ and $\beta = -4\pi \times 6$, as shown in Fig.~\ref{fig:MRbeta}. For each value of $\beta$, we analyze both the GR-like scalarized case ($\xi = 0$) and the teleparallel case with $\xi = 2.5$. We find that the degree of scalarization is governed by the combined effects of the couplings $\beta$ and $\xi$. In particular, scalarization is enhanced for increasing positive values of the teleparallel coupling $\xi$, while it is suppressed for smaller magnitudes of the matter--scalar larger $|\beta|$. For negative values of $\xi$, the opposite trend is expected.

For fixed $\xi$, increasingly negative values of $\beta$ shift the onset of scalarization to higher central densities, indicating a later appearance of scalarized solutions along the stellar sequence and larger deviations from the GR branch, as reflected in both the mass enhancement and the scalar charge.
 Conversely, for fixed $\beta$, increasing $\xi$ amplifies the scalarized regime and steepens the associated mass bump. These results indicate that the matter coupling $\beta$ primarily governs the triggering of scalarization, while the teleparallel derivative interaction controlled by $\xi$ modulates its growth and saturation. The scalarization rate therefore arises from a nontrivial interplay between the two couplings, with $\beta$ setting the instability threshold and $\xi$ regulating the magnitude and persistence of scalar hair.

\subsection{Moment of inertia and observational implications.}
\begin{figure}[h]
	\centering
	\includegraphics[width=0.48\textwidth]{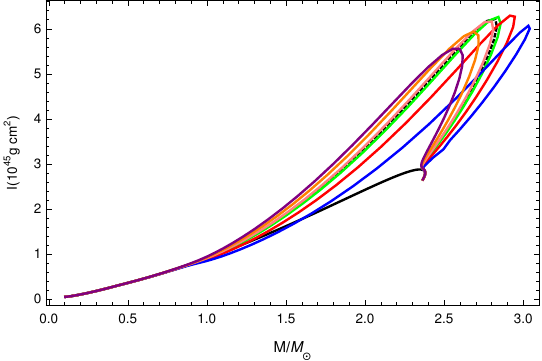}
	\includegraphics[width=0.485\textwidth]{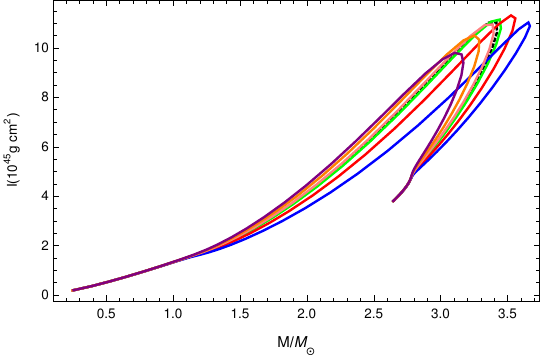}
	\caption{The moment of inertia as a function of the mass  for APR EOS (left) and MS1 EOS (right) using the same parameters of Fig.\ref{fig:plots_apr}}
	\label{fig:inertia}
\end{figure}

The moment of inertia--mass relation provides an additional diagnostic of teleparallel scalarization. In Fig.~\ref{fig:inertia}, we illustrate the variation of the moment of inertia as a function of the gravitational mass. Along the stable branch, configurations with negative values of $\xi$ exhibit systematically smaller moments of inertia than the GR scalarized branch ($\xi=0$), while positive values of $\xi$ lead to enhanced moments of inertia. For identical coupling parameters $(\beta,\xi)$, the stiffer MS1 equation of state yields systematically larger moments of inertia than the softer APR equation of state, reflecting the larger stellar radii and lower compactness of MS1 configurations.

These results have direct observational relevance. NICER measurements of neutron-star masses and radii \cite{Miller:2019,Riley:2019,Miller:2021} constrain stellar compactness and thus indirectly bound the allowed moment of inertia. Even more stringent tests are expected from a future measurement of the moment of inertia in the double-pulsar system PSR~J0737$-$3039A \cite{Morrison:2004,Lattimer:2005,Landry:2018}. The hierarchy
$
I_{\xi<0} < I_{\xi=0} < I_{\xi>0}
$
identified here implies that a $\sim10\%$ determination of $I(M)$ could place meaningful constraints on the sign and magnitude of the teleparallel coupling $\xi$, in analogy with previous studies in scalar--tensor gravity \cite{Doneva:2013,Silva:2015}.

\section{Conclusion}\label{sec:conclusion}

In this work, we have investigated spontaneous scalarization of neutron stars in a teleparallel gravity framework involving a scalar field coupled to both matter and torsion. In the Einstein frame, the theory is characterized by a conformal coupling between matter and the tetrad of the form
\begin{equation*}
A(\phi)=\exp\!\left(\frac{\beta\,\phi^{2}}{2}\right),
\end{equation*}
together with a \emph{derivative torsional interaction} in the action given by
\begin{equation*}
S_{\rm int}=\int d^{4}x\,\sqrt{-g}\;\xi\,\phi^{2}\nabla_{\mu}\phi\,T^{\mu},
\end{equation*}
where $T^{\mu}$ denotes the torsion vector, while $\beta$ and $\xi$ are the corresponding coupling parameters. By constructing static and slowly rotating neutron-star solutions for realistic equations of state, we have examined how the interplay between $\beta$ and $\xi$ modifies the structure of compact stars relative to general relativity and to standard scalar--tensor scenarios.

Our analysis shows that scalarization occurs only within a finite range of central densities and manifests itself through localized deviations in the mass--radius and mass--central density relations. The existence, location, and magnitude of these deviations depend sensitively on the equation of state and on the couplings $(\beta,\xi)$. In particular, the matter coupling controlled by $\beta$ governs the onset of scalarization, while the derivative torsional coupling controlled by $\xi$ modulates the strength and extent of the scalarized regime. We find that positive values of $\xi$ tend to enhance scalarization relative to the general-relativistic scalarized branch, whereas negative values of $\xi$ suppress it. At sufficiently high compactness, scalarization is progressively quenched and the solutions approach the general-relativistic branch.
Moreover, scalarization is found to be bounded: beyond threshold values $\xi_{\min}$ and $\xi_{\max}$, the scalarized solutions saturate and no longer exhibit significant dependence on $\xi$. The behavior of the normalized scalar charge supports this picture, displaying trends consistent with an intermediate scalarized regime that is suppressed at large central densities. In the absence of matter coupling ($\beta=0$), no scalarized solutions are obtained even for nonvanishing $\xi$, indicating that scalarization remains fundamentally matter driven, with the derivative torsional interaction acting as a regulator of its growth and termination.

For slowly rotating configurations, the moment of inertia exhibits a systematic dependence on both the equation of state and the couplings $(\beta,\xi)$, providing an additional probe of teleparallel scalarization effects. Overall, our results suggest that neutron stars provide a useful laboratory for testing teleparallel gravity models with derivative torsional couplings in the strong-field regime. While the present study is restricted to equilibrium configurations, future work incorporating dynamical stability analyses, universal relations, and additional observational channels may further clarify the phenomenological viability of these scenarios and their distinguishability from general relativity and standard scalar--tensor theories.

\newpage
\appendix
\section*{Appendix: Coefficients of structure equations}\label{apdx:coefficients}
\addcontentsline{toc}{section}{Appendix}

The coefficients $ H_{i} $ and $ \Phi_{i} $ of equations (\ref{eq:tov2}) and (\ref{eq:tov3}) are given by
\begin{eqnarray}
	H_{1}(r)&=&-8 \pi  r \phi^{2}\phi'(h-1)^{2}+8 \pi r^{2}\phi\left[\alpha A^{4}h^{2}\phi(3P-\rho)+2\phi'^{2}\right]+32\pi^{2}A^{4}h^{2}\phi^{2}\phi'(\rho-P)  \, , \\
	H_{2}(r)&=&-4\pi \phi^{4}\left(2h^{3}-h^{2}+2h-3\right)+16\pi^{2}r^{2}\phi^{4}\left[-4A^{4}h^{3}P+2A^{4}h^{2}(5P+2\rho)+(2h+3)\phi'^{2}\right] \nonumber \\ && +32\pi^{2}r^{3}\phi^{3}\phi'\left[3\alpha A^{4}h^{2}\phi (3P-\rho)+2\phi'^{2}\right] \, , \\
	H_{3}(r)&=&-32\pi^{2}r \phi^{6}\phi'(h^{2}-4h+3)+256 \pi^{3} r^{3} A^{4} h^{2} \phi^{6}\phi'(\rho+P)+256\pi^{3}r^{4}\alpha A^{4} h^{2}\phi^{6}\phi'(3P-\rho) \, ,\\
	H_{4}(r)&=&256 \pi ^3 r^2 h  \phi ^8 \phi '^2 \, , \\
	\Phi_{1}(r)&=& h\phi^{2}\left(-1+h(2-h)\right)-4\pi r^{2}h\phi^{2}\left(2 A^{4}h^{2}P+A^{4}h(\rho-5P)-\phi'^{2}\right) \nonumber \\ && +4\pi r^{3}\phi \phi'\left(3\alpha A^{4}h^{2}(3P-\rho)\phi+2\phi'^{2}\right) \,  , \\
	\Phi_{2}(r)&=&-4\pi r \phi^{4}\phi'(h^{2}+2h+3)-24\pi r^{2}\phi^{3}\phi'^{2} \nonumber \\ && +32\pi^{2}r^{3}A^{4}h^{2}\phi^{4}\phi'(\rho+P)+32\pi^{2}r^{4}\alpha A^{4}h^{2}\phi^{4}\phi'^{2}(3P-\rho) \, , \\ 
	\Phi_{3}(r)&=&32\pi^{2}r^{2}h\phi^{6}\phi'^{2} \, .
\end{eqnarray}
The coefficient $ C $ is expressed in $ r $-power series form as $ C=\sum_{i=1}^{4} \bar{C}_{i} r^{i}$ where
\begin{align}
	\bar{C}_{1}=&-4-4\pi \xi^{2} \phi^{4} (12+h-2h^{2}+h^{3}) \, \\
	\bar{C}_{2}=&-4\pi \xi \phi' \phi^{2}\left[7+84\pi \xi^{2}\phi^{4}+h^{2}(1+4\pi \xi^{2}\phi^{4})-2h (1+8\pi \xi^{2}\phi^{4})\right] \, \\
	\bar{C}_{3}=&4 \pi  \big[A^4 h^2 \left(-8 \pi \xi ^2  (h-4) P \phi ^4+8 \pi  \xi ^2 \rho  \phi ^4+P+\rho \right)+\xi  \alpha A^4 h^2
	\phi ^2 (3 P-\rho )  \nonumber \\
	&+\phi '^2 \left(32 \pi ^2 \xi ^4 h  \phi ^8+4 \pi \xi ^2 (h+3)  \phi
	^4+2 \xi  \phi +1\right)\big] \\
	\bar{C}_{4}=&16 \pi ^2 \xi  \phi ^2 \phi ' \left[A^4 h^2 \left(8 \pi  \xi ^2 \rho  \phi ^4+8 \pi  \xi ^2 P \phi ^4-P+\rho
	\right)+3\xi \alpha A^4 h^2   \phi ^2 (3 P-\rho )+2 \xi  \phi  \phi '^2\right]
\end{align}

\bibliographystyle{utphys}
\bibliography{refs}

\end{document}